\documentclass[]{article}
\usepackage{amsmath,amssymb,color}
\usepackage{slashed}
\usepackage{bbm}
\usepackage{color}
\usepackage{setspace}
\usepackage{color}
\usepackage{amsmath}
\usepackage{amsfonts}
\usepackage{verbatim}
\usepackage{amssymb}
\usepackage{graphicx}
\usepackage{amsmath,amssymb,calc}
\usepackage[bulletsep]{collref}
\usepackage{bbm}
\usepackage{setspace}
\usepackage{color}
\usepackage{graphicx}
\usepackage{array}
\usepackage{caption}
\usepackage{subcaption}
\newcommand{\be}{\begin{equation}}
\newcommand{\eqb}{\begin{eqnarray}}
\newcommand{\eqf}{\end{eqnarray}}
\newcommand{\bb}{\begin{equation}}
\newcommand{\ee}{\end{equation}}
\newcommand{\beq}{\begin{equation}}
\newcommand{\eeq}{\end{equation}}
\newcommand{\bea}{\begin{eqnarray}}
\newcommand{\eea}{\end{eqnarray}}

\def\lbldef#1#2{\expandafter\gdef\csname #1\endcsname {#2}}

\def\href#1#2{#2}

\newcommand{\ber}{\begin{eqnarray}}
\newcommand{\eer}{\end{eqnarray}}

\newcommand{\beqar}{\begin{eqnarray}}

\newcommand{\eeqar}{\end{eqnarray}}

\newcommand{\eeqarr}{\end{eqnarray}}
\newcommand{\ZZ}{{\rm \kern 0.275em Z \kern -0.92em Z}\;}

\def\CC{{\mathchoice
{\rm C\mkern-8mu\vrule height1.45ex depth-.05ex
width.05em\mkern9mu\kern-.05em}
{\rm C\mkern-8mu\vrule height1.45ex depth-.05ex
width.05em\mkern9mu\kern-.05em}
{\rm C\mkern-8mu\vrule height1ex depth-.07ex
width.035em\mkern9mu\kern-.035em}
{\rm C\mkern-8mu\vrule height.65ex depth-.1ex
width.025em\mkern8mu\kern-.025em}}}

\def\RR{{\rm I\kern-1.6pt {\rm R}}}

\def\ZZ{{\rm Z}\kern-3.8pt {\rm Z} \kern2pt}
\def\IB{\relax{\rm I\kern-.18em B}}
\def\ID{\relax{\rm I\kern-.18em D}}
\def\II{\relax{\rm I\kern-.18em I}}
\def\IP{\relax{\rm I\kern-.18em P}}

\newcommand{\bear}{\begin{eqnarray}}
\newcommand{\eear}{\end{eqnarray}}

\def\to{\rightarrow}

\def\to{\rightarrow}

\title{Chern-Simons-Higgs Theory with Visible and Hidden Sectors and its ${\cal N}=2$ SUSY Extension}
\author{Paola Arias$\,^a$,  Edwin Ireson$^{\,b}$, Fidel A. Schaposnik$\,^c$ and Gianni Tallarita$\,^{d,a}$
~\\
~\\
{$^a\,$\it Departamento de $F\acute{\i}sica$, Universidad de Santiago de Chile} \\{\it Casilla 307, Santiago, Chile}
\\
~\vspace{-3 mm}
\\
{$^b\,$\it Department of Physics, Swansea University}\\
{\it Singleton Park, Swansea, SA2 8PP, UK}
\\
~\vspace{-3 mm}
\\
{$^c\,$\it Departamento de $F\acute{\i}sica$, Universidad Nacional de La Plata/IFLP/CICBA} \\{\it CC 67, 1900 La Plata, Argentina}
\\
~\vspace{-3 mm}
\\
{$^d\,$\it  Centro de Estudios $Cient\acute{\i}ficos$ (CECs),
Casilla 1469, Valdivia, Chile}
 }
\begin{document}
\date{}
\maketitle

\begin{abstract}
We study   vortex  solutions in Abelian Chern-Simons-Higgs theories with visible and hidden sectors.
We first consider the case in which the two sectors are connected through a BF-like gauge mixing term with no explicit  interaction between the the two scalars.  Since first order Bogomolny  equations do not exist in this case, we derive the second order field equations. We then proceed to an ${\cal N}=2$ supersymmetric extension  including a Higgs portal mixing among the visible and hidden charged scalars. As expected, Bogomolnyi equations do exist in this case and we study their string-like solutions numerically.
\end{abstract}
The possible existence of unobserved particles in a hidden sector coupled to the Standard model particles (in the visible sector) or its supersymmetric extensions has recently received much attention in connection
  with dark matter, supersymmetry breaking  and also in the context of phenomenological superstring studies (see \cite{JR}-\cite{Rev} and references therein).

Concerning gauge theories coupled to Higgs scalars there are two possible renormalizable and gauge invariant couplings between the visible and hidden sectors: the gauge kinetic mixing (GKM) and the Higgs portal (HP), originally introduced in refs. \cite{Okun}-\cite{DKM}. Concerning the GKM  coupling, in the case of an Abelian gauge symmetry $U(1)$  and visible and hidden coupled Maxwell-Higgs models, the mixing term $L_{GKM}$ in the  Lagrangian reads
\be
L_{GKM} = \frac12 \xi F_{\mu\nu}G^{\mu\nu}
\label{suno}
\ee
where
\bea
F_{\mu\nu} &=& \partial_\mu A_\nu - \partial_\nu A_\mu\nonumber\\
G_{\mu\nu} &=& \partial_\mu G_\nu - \partial_\nu G_\mu\nonumber
\eea
with $A_\mu$ and $G_\mu$ the visible and hidden gauge fields respectively and $\xi < 10^{-3}$ since larger values are experimentally ruled out \cite{JR},\cite{Rev}.

In the case of the Higgs portal coupling, when the visible $\phi$ and hidden $\eta$ scalars are both taken  as complex charged fields, the mixing term  takes the form
\be
 L_{HP}  = -\frac12 \lambda_{\phi\eta} |\phi|^2 |\eta|^2
\ee
Concerning the portal coupling $\lambda_{\phi\eta}$, for the case in which $\phi$ is the standard model's scalar doublet it is often taken as ${\cal O}(1)$ when it is motivated by baryogenesis or naturalness \cite{Craig} while in the proposal in which $\eta$ introduces a massless Goldstone boson  in connection with the fractional cosmic neutrinos' problem     $\lambda_{\phi\eta} \sim 0.005$ \cite{W}.

 Interestingly enough,  if one wishes to construct a supersymmetric extension in  models with visible and hidden Maxwell-Higgs sectors, one should necessarily have to include both kinetic gauge mixing  and   Higgs portal couplings. Indeed, the  mixing of the
two auxiliary fields $D$, $D'$  belonging to the gauge superfield multiplets
forces a mixing of the scalars in the chiral superfields.
One also needs to add Fayet-Iliopoulos terms in such theories,
to set up, via the Higgs mechanism, the required spontaneous
breaking of
the gauge groups. As a result   a  relation between the mixing parameters $\xi$ and $\Lambda_{\phi\eta}$ is imposed. A discussion of new physics arising from such mixing
can be found in \cite{M2} and references therein.

Now, we know that there is tight connection between supersymmetry and self-dual (Bogomolny or BPS) equations whose solutions also solve  the classical field equations of gauge field theories (see \cite{SY} and references there). In the case of supersymmetric models with visible and hidden Maxwell-Higgs sectors this connection was exploited to find first order BPS equations which exhibit  topologically stable string-like solutions \cite{PS}-\cite{AINS}.

Working in $2+1$ dimensions the  Chern-Simons-Higgs action  also exhibits  vortex-like  solutions  which one can find from self-dual equations when the symmetry breaking is achieved with sixth-order symmetry breaking potential \cite{H}-\cite{JW}. As in the Maxwell action case, such a specific potential is precisely the one required for the supersymmetric extension of the model  \cite{LLW}.

It is the purpose of the present work to study $2 + 1$ dimensional models with visible and hidden sectors in which the gauge field dynamics is dictated by Chern-Simons (CS)  actions. The natural gauge-field mixing in this case takes the form of a $BF$-term \cite{BBRT}
\be
L_{GM} = \frac12 \xi \varepsilon^{\alpha\beta\gamma}A_\alpha G_{\beta\gamma}
\label{GM}
\ee
It should be noticed that apart from the interest {\it per se} of the $BF$ theories, this kind of term has played a relevant role in the study of self-duality in topologically massive gauge theories \cite{vN}-\cite{DJ} and $2+1$ bosonization \cite{FS}-\cite{LMNS} with interesting applications in condensed matter problems \cite{F1}-\cite{F2}.

We shall first consider the case in which visible and hidden sectors are coupled solely by the gauge field mixing \eqref{GM} and then discuss the supersymmetric extension. In this last case  a Higgs portal is necessarily present and  consists of two sixth order terms of the form $(|\phi|^2)^2 |\eta|^2$ and $|\phi|^2  (|\eta|^2)^2$ with mixing coupling constants related to the gauge field parameters in such a way as to ensure supersymmetry and, a fortiori, the existence of BPS equations.

~

\noindent{\bf A Chern-Simons-Higgs model with visible and hidden sectors}

We start from the Lagrangian for   $2+1$ dimensional  with   visible and hidden $U(1)$ gauge fields and dynamics governed by Chern-Simons terms coupled to two charged scalars and a gauge mixing term of the form \eqref{suno},
\be
\mathcal L=  \frac14 \kappa \varepsilon^{\alpha\beta\gamma}A_\alpha F_{\beta\gamma} + \frac14 \kappa_h \varepsilon^{\alpha\beta\gamma}G_\alpha G_{\beta\gamma}  + \frac12 \xi \varepsilon^{\alpha\beta\gamma}A_\alpha G_{\beta\gamma}  + (D_\mu\phi)^*(D_\mu ^A\phi)  +(D_\mu\eta)^*(D_\mu^G\eta) - V[\phi]- V[\eta]
\label{dosuno}
\ee
where the covariant derivatives are defined as
\be
D_\mu^A\phi = \partial_\mu \phi - ie A_\mu\phi \; , \;\;\;\;\;
D_\mu^G\eta = \partial_\mu \eta - ig G_\mu\eta
\ee
The symmetry breaking potentials $V[\phi]$ and $V[\eta ]$ will be fixed below. The Chern-Simons couplings $\kappa,\kappa_h$ and $\xi$ are real constants with dimensions of a mass and   Minkowski metric is taken as $\,g_{\mu\nu}=$  diag$(1,-1,-1)$.
Clearly, in the limit of $\xi \rightarrow 0$ we have two decoupled Chern-Simons-Higgs theories which, for a particular choice of symmetry breaking potential have shown to be self-dual \cite{H}-\cite{JW}, this being connected with the possibility of an ${\cal N}=2$  supersymmetric extension \cite{LLW}.

The  gauge field equations read
\be
\frac{\kappa}2 \varepsilon^{\mu\alpha\beta} F_{\alpha\beta} +
\frac{\xi}2  \varepsilon^{\mu\alpha\beta} G_{\alpha\beta} = J^\mu [\phi]
\;, \;\;\;\;\;\;\;\;
\frac{\kappa_h}2 \varepsilon^{\mu\alpha\beta} G_{\alpha\beta} +
\frac{\xi}2  \varepsilon^{\mu\alpha\beta} F_{\alpha\beta} = J^\mu [\eta],
\ee
with the conserved matter currents $J^\mu (\phi), J^\mu (\phi)$ given by
\be
J_\mu[\phi] = {ie}(\phi^* D_\mu^A \phi - \phi (D_\mu^A \phi)^*) \;, \;\;\;\; \;\;\;\;
J_\mu [\eta]= {ig}(\eta^* D_\mu^G \eta - \eta (D_\mu^G \eta)^*)
\ee

The time components of Eqs. \eqref{ecA}-\eqref{ecG}  read
\be
 - \kappa B_A - \xi B_G= J^0[\phi] \; , \;\;\;\; \;\;\;\;
  -\kappa_h B_G -\xi B_A = J^0[\eta] \label{mag2}
\ee
Magnetic and electric fields are defined as
\be
B_A = -F^{12} \;, \;\;\;\; \;\;\;\; B_A = -G^{12} \;, \;\;\;\; \;\;\;\;  E_A^i = F^{0i} \;, \;\;\;\; \;\;\;\; E_B^i = G^{0i}
\ee

Concerning the field equations for the visible and hidden charged scalars, they read
\be
D_{\mu}(A) D^{\mu} (A)\phi = -\frac{\partial V(\phi)}{\partial\phi^*} \;, \;\;\;\;\;\;\;\;
D_{\mu}(G) D^{\mu}(G) \eta = -\frac{\partial V(\eta)}{\partial\eta^*}
\ee
 As already mentioned, when there is just one sector and the Higgs potential is  sixth order  the CS-Higgs model becomes self-dual and one finds first order field equations whose static rotationally symmetric solutions also solve the Euler-Lagrange equations. This can be done \`a la Bogomolny \cite{bogo}, by writing the energy as a sum of squares plus a topological term  \cite{H}-\cite{JW}.
In the present case,  because of the gauge field mixing  the energy  cannot be written in this way so that  first order Bogomolny equations do not exist for any choice of Higgs potentials in which the two scalars do not interact. Of course,
one can solve the second order field equations \eqref{dosuno} with vortex like solutions.
We shall now discuss this issue choosing sixth-order potentials for both scalars, {which is the standard choice for the system with one sector only leading to BPS equations}  \cite{H}-\cite{JW},
\be
V[\phi] = \frac{e^4}{4\kappa^2}|\phi|^2 (|\phi|^2 - \phi_0^2)^2  \;, \;\;\;\;\;\;\;\;
V[\phi] = \frac{g^4}{4\kappa^2_h}|\eta|^2 (|\eta|^2 - \eta_0^2)^2.
\ee
It is convenient to work with dimensional variables
$
x \rightarrow x/(e\phi_0^2),    \phi \rightarrow \phi \phi_0, \eta \rightarrow \eta \phi_0, A_\mu \rightarrow A_\mu \phi_0^2, G_\mu \rightarrow G_\mu \phi_0^2$ and redefine coupling constants
  $h\equiv g/e$ and $k=\kappa/e$,    $k_h=\kappa_h/g$, $\bar \xi=\xi/g$, $\bar\eta_0=\eta_0/\phi_0$.

Then, after proposing the usual Nielsen-Olesen \cite{NO} ansatz, which in polar coordinates $(t,\theta)$ reads
\bea
&& \phi= f(r) e^{in_1\theta},\,\,\,\,\,\,\, A_\theta=-\frac{ a(r)}r, \,\,\,\,\,\, A_r=0 \nonumber\\
&& \eta= q(r) e^{in_2\theta},\,\,\,\,\,\,\, G_\theta=-\frac{ b(r)}{r}, \,\,\,\,\,\, G_r=0
\label{ansatz}
\eea
the resulting field equations read
\bea
&& 2A_0 f^2={k} \frac{a'}r +h{\bar \xi} \,\frac{b'}r, \\
&& 2hG_0 q^2={k_h} \frac{b'}r +{\bar \xi}\, \frac{a'}r,\\
&& \frac{2}r \left(-n_1+a\right) f^2= {k} A'_0+h{\bar \xi}G_0',\\
&& \frac{2}r \left(-n_2+h b\right) q^2= {k_{h}} G'_0+{\bar \xi}A_0',\\
&& f''+\frac{f'}{r}- \frac{1}{r^2}\left(-n_1+a\right)^2 f+A_0^2 f= \frac{1}{4k^2} f \left(4f^2-3f^4-1\right),\\
&& q''+\frac{q'}{r}- \frac{1}{r^2}\left(-n_2+h b\right)^2 q+h^2G_0^2 q= \frac{h^2}{4k_h^2} q \left(4  q^2-3q^4-\tilde \eta_0^4\right).
\eea
{\indent The vortex-like solutions of these equations, in which both scalar fields condense in the vacuum breaking their respective gauge symmetries, are qualitatively similar to those found in an analogous supersymmetric model which we investigate below. In particular, the results of variations of the gauge mixing parameter on the field profiles are identical in both systems. Therefore, in order not to present similar plots repeatedly, we refer the reader to the supersymmetric section for numerical solutions.} \newline
At this point it is interesting to note that when just one of the $U(1)$ symmetries is spontaneously broken the theory can be cast in a very simple way. Let us assume that the {hidden} sector has an unbroken $U(1)$ symmetry, while the visible one has a broken one (we could have
chosen the other way around as well). The simplest way to achieve this is by eliminating
the hidden scalar sector so that all $\eta$ dependent terms    in Lagrangian
\eqref{dosuno} are absent. In that case
the kinetic mixing term can be eliminated by rotating the hidden field through the term
\be
\tilde G_\mu=G_\mu+\frac{\xi}{\kappa_h} A_\mu,
\label{rot}
\ee
so that the  Lagrangian reduces to one with no gauge fields mixing,
\be
L =  \frac14 \kappa_h \varepsilon^{\alpha\beta\gamma}\tilde G_\alpha \tilde G_{\beta\gamma}   + \frac14 \tilde \kappa \varepsilon^{\alpha\beta\gamma}A_\alpha F_{\beta\gamma}  + (D_\mu[A]\phi)^*(D_\mu[A]\phi)  - V[\phi].
\ee
where $\tilde \kappa \equiv \kappa- {\xi^2}/{\kappa_h}$.  The equations of motion for $A_\mu$ then become the ones  of the ordinary CS-Higgs model,
\be
\frac{\tilde \kappa}2 \varepsilon^{\mu\alpha\beta} F_{\alpha\beta}  =
{ie}(\phi^* D^\mu \phi - \phi (D^\mu \phi)^*)
\label{ecA}
\ee
while that for  $\tilde G_\mu$ corresponds to the pure CS theory,
\be
\frac{\kappa_h}2 \varepsilon^{\mu\alpha\beta} \tilde G_{\alpha\beta}  = 0 \, \,\, \Rightarrow
\frac{\kappa_h}2 \varepsilon^{\mu\alpha\beta} \tilde G_{\alpha\beta} = -{\xi}
\varepsilon^{\mu\alpha\beta} \tilde F_{\alpha\beta}
\label{ecG}
\ee
Using this equation and eq.(\ref{rot}) one finds a relation between the original hidden and visible magnetic fields
\be
B_G=-\frac{\xi}{\kappa_h}B_A.
\ee
Hence, although the hidden gauge symmetry is unbroken the existence of vortices with magnetic flux in the sector where the symmetry is broken implies a  non-trivial magnetic flux $\Phi_G$ for the hidden sector, due to
gauge mixing of the sectors. {In particular, the vortex like solution for the magnetic field profile in the broken symmetry sector implies through this relation a vortex like magnetic field profile for the unbroken sector.}

The time component of eq.(\ref{ecA}) allows us to write the   usual Gauss-like relations of CS theories between
$A_0$ and the magnetic field, namely
\be
A_0=\tilde \kappa \frac{B_A}{2e^2|\phi|^2}
\ee
Inserting this relation in the $T_{00}$ component of the energy-momentum tensor

we find for the energy
\be
E = \int d^2x \left( |\nabla\phi - ie A_i\phi|^2
+ \frac{\kappa^2B_A^2}{4e^2|\phi|^2} \left(1 - \frac{\xi^2}{\kappa\kappa_h}
\right) ^2+ V[\phi]
\right)
\label{energyef}
\ee
which is just the expression for the energy of the ordinary  self-dual  Chern-Simons vortices discussed in \cite{H}-\cite{JW} if the Chern-Simons $\kappa$ parameter is taken as
\be
  \kappa = \kappa\left(1 - \frac{\xi^2}{\kappa\kappa_h}
\right)
\ee
so that the usual magnetic flux/matter charge  can be seen to be, in the visible sector
 \be
 Q_{vis} = \int d^2x J^0= - \kappa   \Phi_A
 \label{29}
 \ee
One can also see the visible sector acts as a source of electric charge $Q_{\rm hid}$ for the hidden one, since taking the zeroth component of eq.\eqref{ecG} and using \eqref{29} one finds that
\be
Q_{\rm hid} = -\frac{\xi}{\kappa_h} Q
\ee

 {Since in the present case the energy of the model  with visible and hidden sector
has been reduced to an effective model with just one sector, one can find a Bogomolny
bound for the energy \eqref{energyef} that is saturated satisfying the usual CS-Higgs
self-dual equations first obtained in refs.\cite{H}-\cite{JW}}

{{In this case, where the system diagonalizes to a single sector, the numerical solutions have} been described in detail in \cite{H}-\cite{JW}
and we shall solely mention here that in contrast to the case of the Nielsen-Olesen vortex, in which the magnetic field has its maximum at its center, the magnetic
field for the Chern-Simons vortex is concentrated in a
ring surrounding the zero of the Higgs field and a similar behavior holds for the electric field.}

~

\noindent{\bf The ${\cal N} = 2$ supersymmetric extension}

Within the  ${\cal N}= 2$ supersymmetric extension of Chern-Simons theories coupled
to charged fields   one introduces ({in a language analogous to  4D $\mathcal{N}=1$ supersymmetry}) a vector superfield $V$,
composed (in the Wess-Zumino gauge)  of a gauge field $A_\mu$, a 2-component complex spinor $\chi$, a real scalar field $M$  and the auxiliary scalar field $D$ \cite{LLW},\cite{Schwarz}-\cite{KT}
\be
V = 2i\theta\bar\theta M + 2\theta\gamma^\mu\bar\theta A_\mu + \sqrt2 i\theta^2 \bar\theta  \bar \chi-\sqrt2 i \bar\theta^2\theta \chi + \theta^2\bar\theta^2 D
\ee
with the $\gamma$-matrices taken as the Pauli matrices with $\gamma^0 = \sigma^3,
\gamma^i = \sigma^i$.

Concerning matter content, one introduces a chiral  superfield $\Phi$  a complex scalar $\phi$, a complex fermion $\sigma$ and an auxiliary field $F$ components.

The Chern-Simons-matter action composed of these superfields reads
\be
S = \int d^3x\int d^2\theta d^2\bar\theta  \left( \kappa V\Sigma +  \Phi^\dagger \exp^{-ieV} \Phi
\right)
\label{ection}
\ee
with  $\Sigma$ (the linear multiplet) defined as
\be
\Sigma = \bar D^\alpha D_\alpha V
\ee
The component fields  Lagrangian derived from the superaction \eqref{ection} takes the form
\bea
& & L=\kappa(\epsilon^{\mu\nu\rho}A_\mu \partial_\nu A_\rho -\bar{\chi}\chi + 2 D M) + |D_\mu \phi|^2
+i \bar{\sigma} \gamma^\mu D_\mu \sigma + |F|^2 - ({e^2}M^2- {e}D)|\phi|^2 - {e\phi_0^2} D  \nonumber\\
& & -{e}\bar{\sigma}\sigma M + i {e}\bar{\chi}\sigma \phi - i {e}\bar{\sigma}\chi \phi.
\label{lagrangianCSH}
\eea
Concerning   SUSY variations, one has
\be
\begin{array}{lll}
 -2i \delta A_\mu=\bar{\epsilon}\gamma_\mu \chi -\bar{\chi}\gamma_\mu \epsilon &
 2\delta M=\bar{\epsilon}\chi -\bar{\chi}\epsilon & \\
  \;\delta \phi= \bar{\epsilon} \sigma & 2\delta D= \bar{\epsilon}\gamma^\mu \partial_\mu \chi- \partial_\mu
\bar{\chi}\gamma^\mu \epsilon & i \delta F= \bar{\epsilon}\gamma^\mu D_\mu \sigma.\nonumber\\
 \;\delta \chi=[{i}\epsilon^{\mu\nu\rho}\gamma_\rho F_{\mu\nu}- i D - \gamma^\mu\partial_\mu M]\epsilon &
\delta \sigma=(-i \gamma^\mu D_\mu \phi-M\phi + F)\epsilon
\end{array}
\label{transfor2}
\ee

The F term vanishes, since we impose no superpotential, and one can eliminate the non-dynamical fields $M$ and $D$ from their algebraic field equations:
\beq
-2\kappa^2 D={e^3}|\phi|^2 (|\phi|^2-\phi_0^2),\;\;\;-2\kappa M={e}(|\phi|^2 -\phi_0).
\eeq
When reintroduced back into the action, this generates the sixth-order potential originally proposed in \cite{H}-\cite{JW} in the search of BPS equations  and derived in the ${\cal N}=2$ supersymmetric  context in \cite{LLW}
\beq
V_{eff}=  {e^4}\frac{|\phi|^2}{4\kappa^2} (|\phi|^2 -\phi_0^2)^2
\eeq

Indeed, within the context of supersymmetry  the first order BPS equations can be obtained by imposing that all
fermions and their supersymmetric variations vanish everywhere \cite{SY},   as first shown for the CS theory in\cite{LLW}.

We are now ready to consider Chern-Simons-Higgs actions both for the visible and for the hidden sectors  coupled through a CS like mixing term.     We then consider as a starting point the supersymmetric action
\begin{equation}
S_{{\cal N}=2}=\kappa S_1[V, \Phi] + \kappa_h S_2 [U, \Omega ]-\xi S_{int}[V,U].
\end{equation}
Here $V, \Phi$ and $U,\Omega$ are the vector and scalar superfields of the visible and hidden sectors respectively. The components of $V$ and $\Phi$ are those introduced in above.  Concerning the hidden vector and chiral scalar superfields, we denote their components as $U= (N,G_\mu, \tau,d)$ and $\Omega=(\eta,\rho,f)$.

The interaction  $S_{int}$  takes the form
\beq
\int d^3x d^4\theta \; U\Sigma = \int d^3x d^4\; \theta V\Upsilon=\int d^3x \;\left(  \epsilon^{\mu\nu\rho}A_\mu G_{\nu\rho}-\frac{1}{2}\left(\bar{\chi}\tau+\bar{\tau}\chi \right) +{2DN+ 2dM} \right)
\label{csportal}
\eeq

From the field equations of auxiliary fields $D,d,M,N$ one gets:

\be
\begin{array}{ll}
  \frac{\delta~}{\delta D}:\text{   }\kappa M={-\frac{e}{2}}(|\phi|^2- \phi_0^2) +\xi N \hspace{0.4 cm}&  \hspace{0.4 cm}
  \frac{\delta~}{\delta d}:\text{  } \kappa_h N={-\frac{g}{2}}(|\eta|^2- \eta_0^2)+ \xi M \\
  ~ & ~
  \\
  \frac{\delta~}{\delta M}:\text{  }  \kappa D = {e^2} M |\phi|^2+{\xi d}
  \hspace{0.4 cm}& \hspace{0.4 cm}
 \frac{\delta~}{\delta N}:\text{  }  \kappa_h d = {g^2} N |\eta|^2+ {\xi D}
 \end{array}
 \label{37}
\ee
These are two 2x2 linear systems of equations, the solution of which is
\bea
  M &=&{-\frac{1}{2(\kappa\kappa_h-\xi^2)}}\left( {\kappa_h e} (|\phi|^2- \phi_0^2)+  {\xi g} (|\eta|^2- \eta_0^2)\right)  \\
  N &=&{-\frac{1}{2(\kappa\kappa_h-\xi^2)}}\left( {\kappa g} (|\eta|^2- \eta_0^2)+ {\xi e} (|\phi|^2- \phi_0^2)\right)  \\
 D &=& {-\frac{{e^2}|\phi|^2}{2(\kappa\kappa_h-\xi^2)(\kappa -  \xi) }\left(\vphantom{\sum} (e\kappa_h|(\phi|^2- \phi_0^2)
  +g\xi (|\eta|^2- \eta_0^2)\right)}  \\
     d &=& {-\frac{{g^2}|\eta|^2}{2(\kappa\kappa_h-\xi^2)(\kappa_h -  \xi) }
     \left(\vphantom{\sum}  e\xi    (|\phi|^2- \phi_0^2) +           {g}\kappa(|\eta|^2- \eta_0^2) \right)}
\eea

Using eqs.\eqref{37} in the Lagrangian and using eqs.\eqref{37} the effective potential  becomes simply
{\be
 V_{eff} =  {\frac{1}{4(\kappa\kappa_h-\xi^2)}}
 \left(\left( {\kappa_h e} (|\phi|^2- \phi_0^2)+  {\xi g} (|\eta|^2- \eta_0^2)\right)^2 e^2|\phi|^2 +
 \left( {\kappa g} (|\eta|^2- \eta_0^2)+ {\xi e} (|\phi|^2- \phi_0^2)\right)^2  g^2|\eta|^2
 \right)
 \ee
 }
and we have the symmetric BPS equations
{
\be
\begin{array}{ll}
  F_{12}=\pm \frac{e^2|\phi|^2}{4(\kappa \kappa_h-\xi^2)(\kappa - \xi)}\left(\vphantom{\frac{\phi^2}{\xi^2}} {e\kappa_h} (|\phi|^2- \phi_0^2)+ {\xi g} (|\eta|^2- \eta_0^2)\right) \hspace{0.4 cm} &
 \hspace{0.4 cm}
  D_1[A]\phi \pm D_2[A] \phi=0  \\
  ~ & ~\\
  G_{12}=\pm\frac{g^2|\eta|^2}{4(\kappa \kappa_h-\xi^2)(\kappa_h - \xi)}\left( \vphantom{\frac{\phi^2}{\xi^2}} {g\kappa} (|\eta|^2- \eta_0^2)+ {e\xi} (|\phi|^2- \phi_0^2)\right) \hspace{0.4 cm} & \hspace{0.4 cm}
   D_1[G]\eta \pm D_2[G] \eta=0
\end{array}
\ee
}
These are the Bogomolny equations of the supersymmetric Lagrangian  \eqref{lagrangianCSH} with all fermionic fields put to zero and auxiliary fields replaced using their algebraic field equations
\be
L=\kappa\epsilon^{\mu\nu\rho}A_\mu \partial_\nu A_\rho  +  \kappa_h\epsilon^{\mu\nu\rho}G_\mu \partial_\nu G_\rho + 2\xi\epsilon^{\mu\nu\rho}A_\mu \partial_\nu G_\rho + |D_\mu[A]\phi|^2 + |D_\mu[G]\eta|^2 - V_\text{eff}[\phi,\eta]
\label{lagrangian}
\ee
Switching to the dimensionless units
$
\rho \rightarrow \phi_0^2 r, \;\; \tilde{\phi} = \phi/\phi_0 ,\;\; \tilde{\eta} = \eta/\eta_0$
and using  a rotational ansatz like that in \eqref{ansatz} the BPS equations reduce to
{\be
\begin{array}{ll}
\frac{1}{\rho}\partial_\rho A_\theta =-\frac{e^2|\tilde{\phi}|^2}{4(\kappa \kappa_h-\xi^2)(\kappa - \xi)}\left(e\kappa_h(|\tilde{\phi}|^2- 1)+ \xi g(|\tilde{\eta}|^2- \frac{\eta_0^2}{\phi_0^2})\right) \hspace{0.4 cm} & \hspace{0.4 cm}  \rho \partial_\rho\tilde{\phi} = \tilde{\phi}\left(1-e A_\theta\right)\\
~ & ~\\
\frac{1}{\rho}\partial_\rho G_\theta = -\frac{|\tilde{\eta}|^2}{4(\kappa \kappa_h-\xi^2)(\kappa_h-\xi)}\left(\xi e(|\tilde{\phi}|^2- 1)+ \kappa_h g(|\tilde{\eta}|^2- \frac{\eta_0^2}{\phi_0^2})\right)
\hspace{0.4 cm} & \hspace{0.4 cm}
 \rho \partial_\rho\tilde{\eta} = \tilde{\eta}\left(1-g G_\theta\right)
 \end{array}
 \label{BPSi}
 \ee
 }
 Concerning the Gauss Law resulting from Lagrangian \eqref{lagrangian} using again eqs.\eqref{37} they can be  written in the form
 \bea
 A_0 &=& \frac{M}2    = {-\frac{1}{4(\kappa\kappa_h-\xi^2)}}\left( {\kappa_h e} (|\phi|^2- \phi_0^2)+  {\xi g} (|\eta|^2- \eta_0^2)\right)\nonumber\\
 G_0 &=&  \frac {N}2 = {-\frac{1}{4(\kappa\kappa_h-\xi^2)}}\left( {\kappa g} (|\eta|^2- \eta_0^2)+ {\xi e} (|\phi|^2- \phi_0^2)\right)
 \eea

The appropriate boundary conditions {for string-like solutions} are
\be
\begin{array}{ll}
\tilde{\phi}(0)=0,\;\; \tilde{\eta}(0)=0 \hspace{0.4 cm} & \hspace{0.4 cm}
A(0) = 0 ,\;\; G(0)=0 \\
\tilde{\phi}(R)=1,\;\; \tilde{\eta}(R)=\eta_0/\phi_0
 \hspace{0.4 cm} & \hspace{0.4 cm}
A(R) = 1/e ,\;\; G(R)=1/g,
\end{array}
\label{konsi}
\ee
where in the second line the limit of the radial variable $R \to \infty$ should be taken. {These solutions are topologically stable}.

In order to solve \eqref{BPSi} we used a finite difference numerical procedure taking  a value $R = 40$ for the large radial distance in \eqref{konsi}. We show in figure 1 the resulting hidden and visible charged scalars, magnetic and electric fields for different values of the hidden gauge coupling $g$ keeping all other parameters fixed. Although the value chosen for the mixing parameter is not very small, we see that the visible fields are not much affected by the changes in $g$. In contrast, as it was to be expected, as the coupling between the hidden gauge field and charge scalar grows the vortex core becomes   tighter and tighter and the magnitude of the magnetic and electric fields maxima get higher.  Due to the symmetry of the BPS equations under interchange of the visible and hidden fields an identical result with their role interchanged {for the visible sector} is obtained when instead of $g$ one considers different values of the visible coupling constant $e$.

We study in figure 2 the behavior of scalar, magnetic and electric fields under changes in the mixing parameter $\xi$. We see that both the magnitude of magnetic and electric fields become larger as $\xi$ grows. Figure 3 shows the dependence of the scalar and magnetic fields under changes in the scalar vevs relative value. Tighter vortex cores and larger magnetic field maximums correspond to larger ratios $\eta_0/\phi_0$. We see that as the scalar vev is lowered, there is a smearing of the electric field towards larger distances.

~

\noindent{\bf{Summary and Conclusions}}

We have studied visible and hidden sectors in  theories in which gauge dynamics is dictated by Chern-Simons actions including the  ${\cal N} = 2$ supersymmetric extension. Instead of the usual kinetic gauge mixing introduced in $3+1$ space-time dimensions,  in the planar case the natural gauge mixing is a BF term. As for the portal term arising in the supersymmetric extension it corresponds to  sixth-order polynomials mixing the two charged scalars.

When the portal term is not included, no Bogomolny first order equations can be found. A particularly interesting phenomenon takes place  in this case when  one of the two charged scalars is absent. Hence there is no gauge symmetry breaking in the corresponding sector but, nevertheless, vortex-like solutions exist with both quantized magnetic flux and electric charge.

Concerning the supersymmetric extension, we found the associated BPS equations and studied them numerically, determining in particular how the magnetic, electric and  Higgs field profiles depend on the relevant parameters constructed from coupling constants and masses of the model.

As stated in the introduction, BF terms associated to CS theories have played a relevant role concerning self-duality in topologically massive theories and bosonization of Thirring like theories in $2+1$ dimensions \cite{FS}-\cite{F2}. We expect to study this issue in the case of models with hidden sectors in a forthcoming publication.

~

\noindent{\bf{Acknowledgments:}}
F.A.S  would like to thank Gustavo Lozano for comments and discussions. P.A. is supported by FONDECYT project 11121403 and Anillo ACT 1102.
E.I. is supported by a STFC studentship.
F.A.S. is financially supported by Pip-CONICET,
PICT-ANPCyT, UNLP and CICBA grants.
G.Tallarita Fondecyt grant 3140122.  CECs center is funded by the Chilean
Government through the Centers of Excellence Base Financing Program of
Conicyt.

\begin{figure}[b]
\begin{subfigure}{.5\textwidth}
\centering
\includegraphics[width=0.8\linewidth]{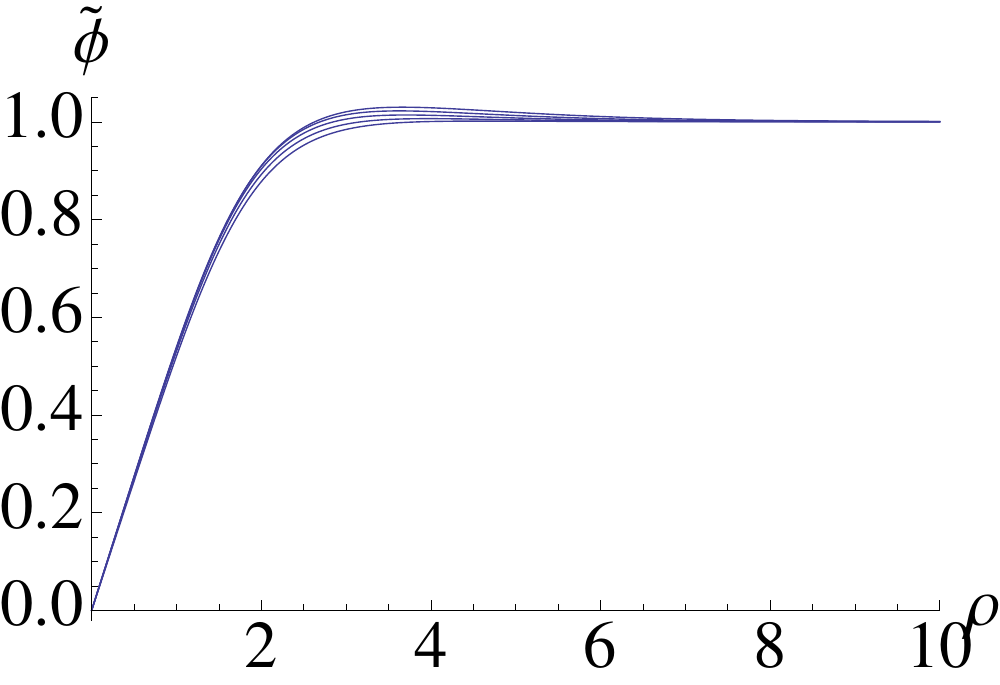}
\end{subfigure}
\begin{subfigure}{.5\textwidth}
\includegraphics[width=0.75\linewidth]{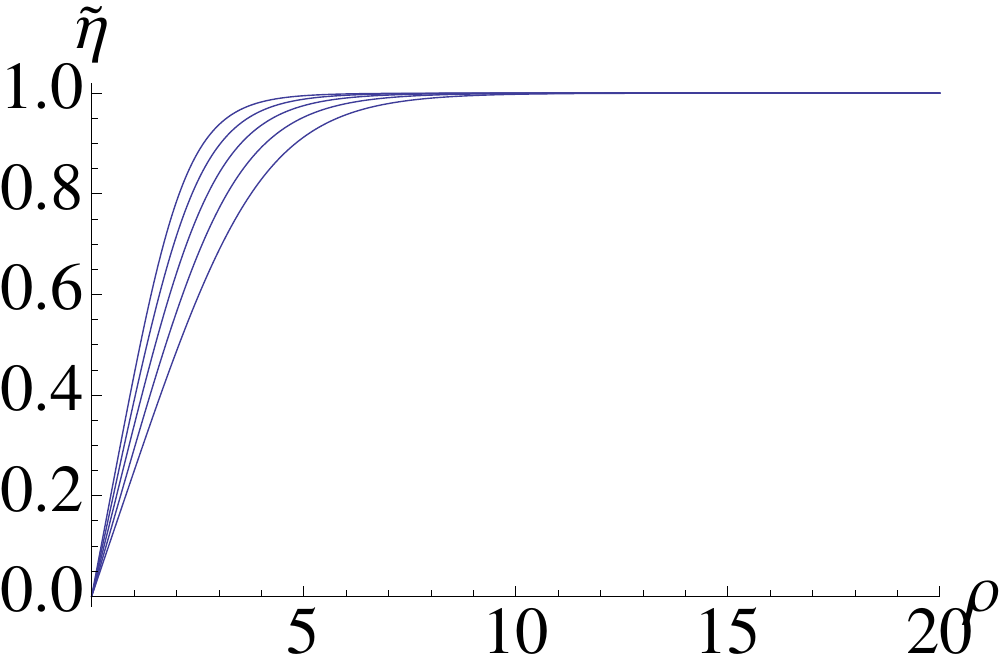}
\end{subfigure}\\
\begin{subfigure}{.5\textwidth}
\centering
\includegraphics[width=0.75\linewidth]{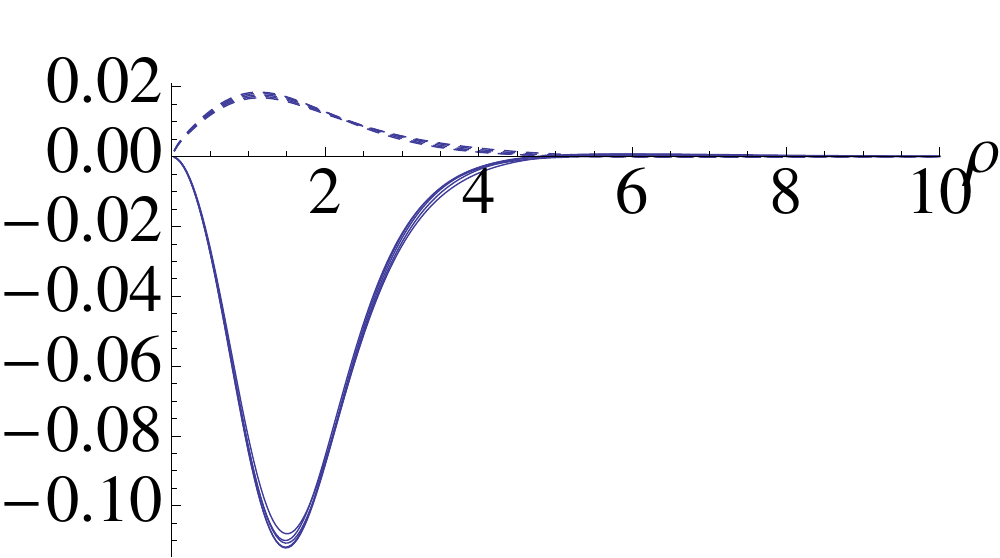}
\end{subfigure}
\begin{subfigure}{.5\textwidth}
\includegraphics[width=0.9\linewidth]{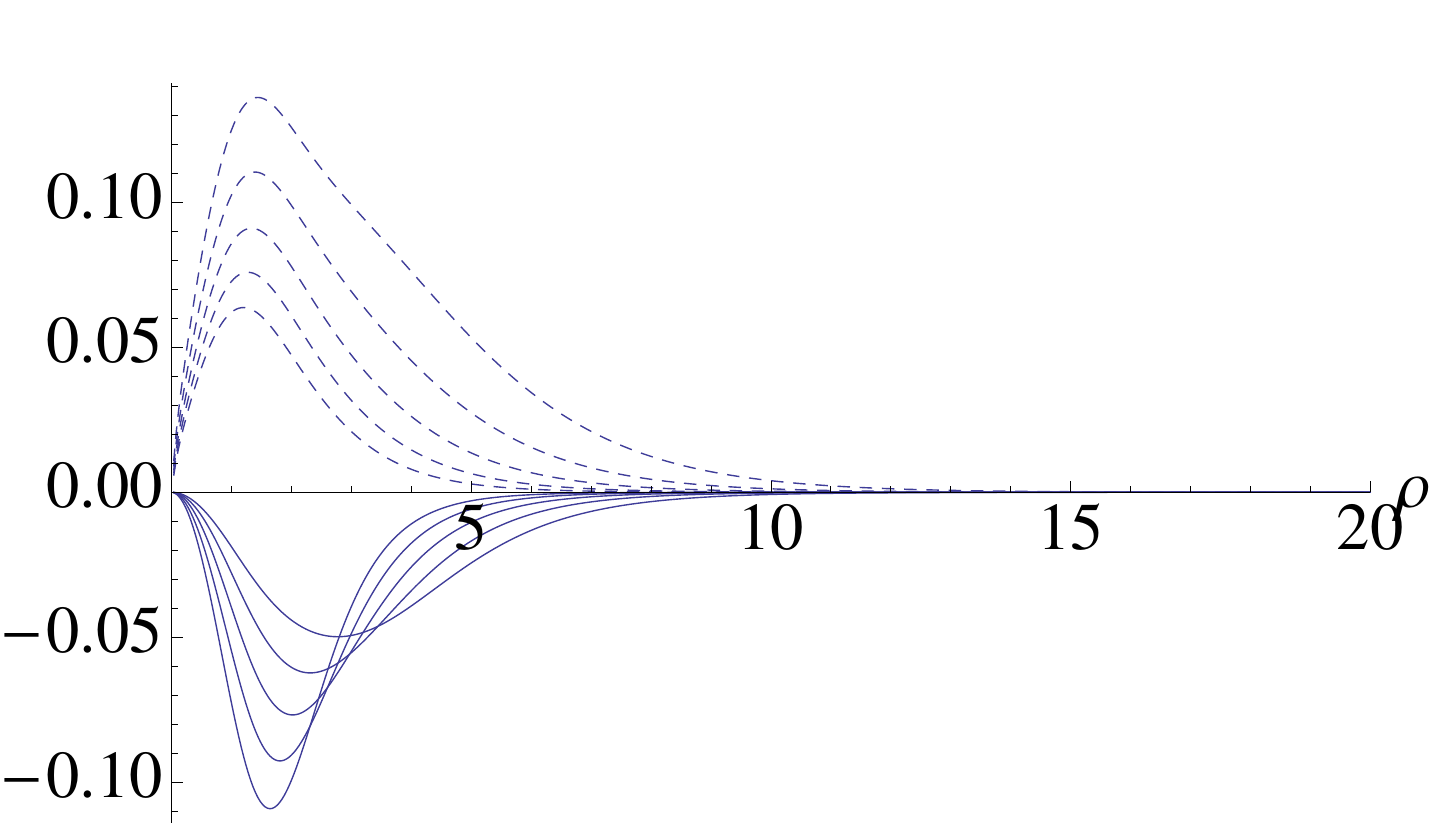}
\end{subfigure}
\caption{Field profiles at $\kappa = 2$, $\kappa_h = 3$, $\phi_0/\eta_0 = 1$, $e = 3$ and $\xi=1$, changing $g$ between $1.4$ and $2.2$ in steps of $0.2$, with highest $g$ corresponding to tighter vortex core. Plots on the second line include electric (dashed line) and magnetic fields (solid line) for both the $A$ (left column) and $G$ (right column) fields. }%
\label{fig4}%
\end{figure}
\begin{figure}[h]
\begin{subfigure}{.5\textwidth}
\centering
\includegraphics[width=0.7\linewidth]{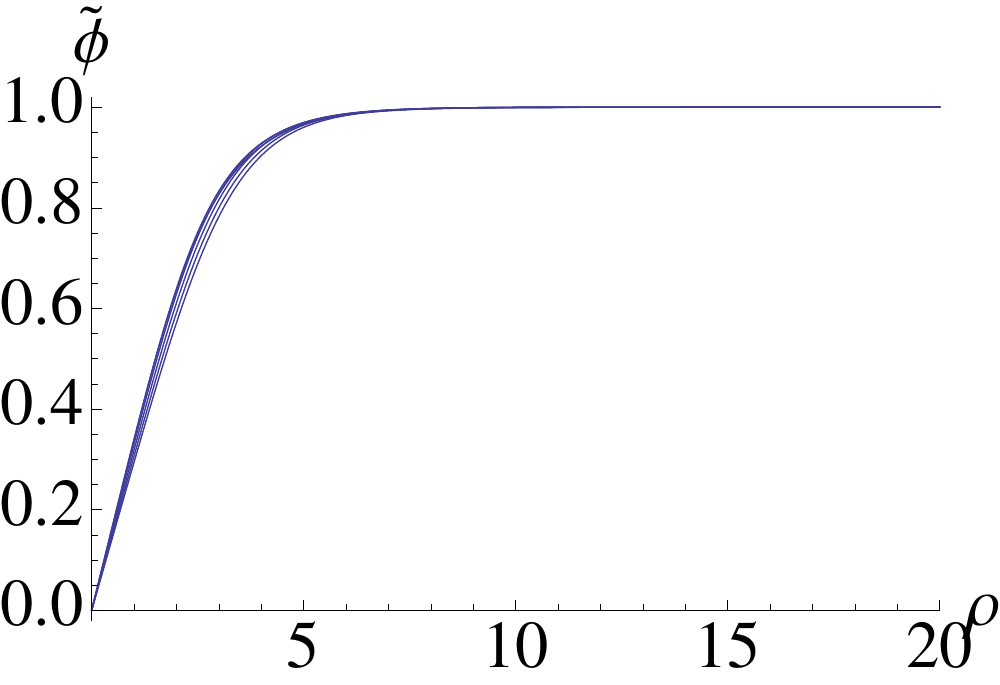}
\end{subfigure}
\begin{subfigure}{.5\textwidth}
\includegraphics[width=0.8\linewidth]{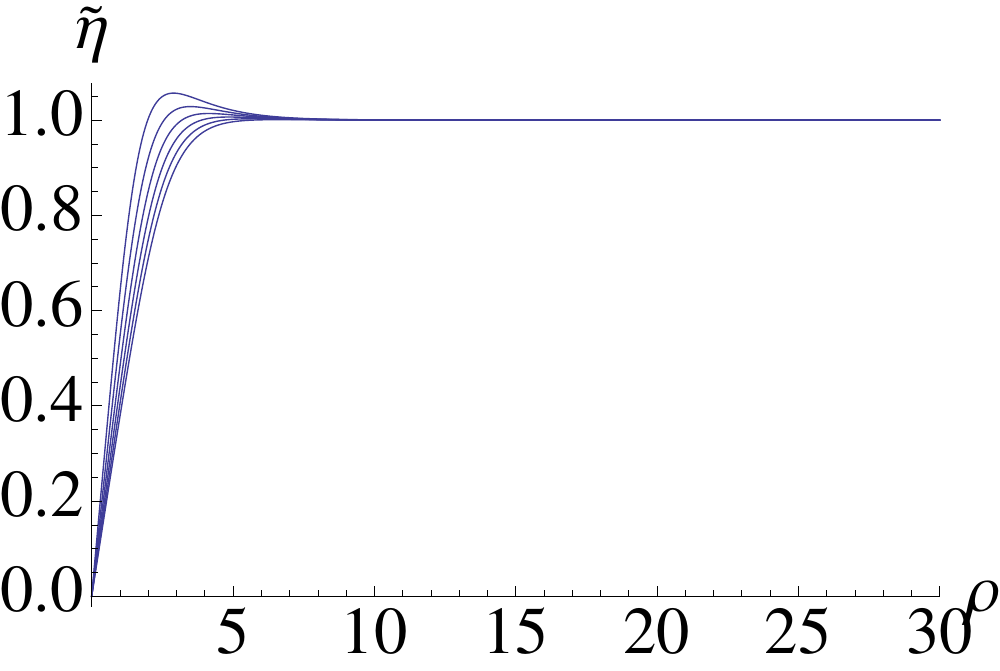}
\end{subfigure}\\
\begin{subfigure}{.5\textwidth}
\centering
\includegraphics[width=0.9\linewidth]{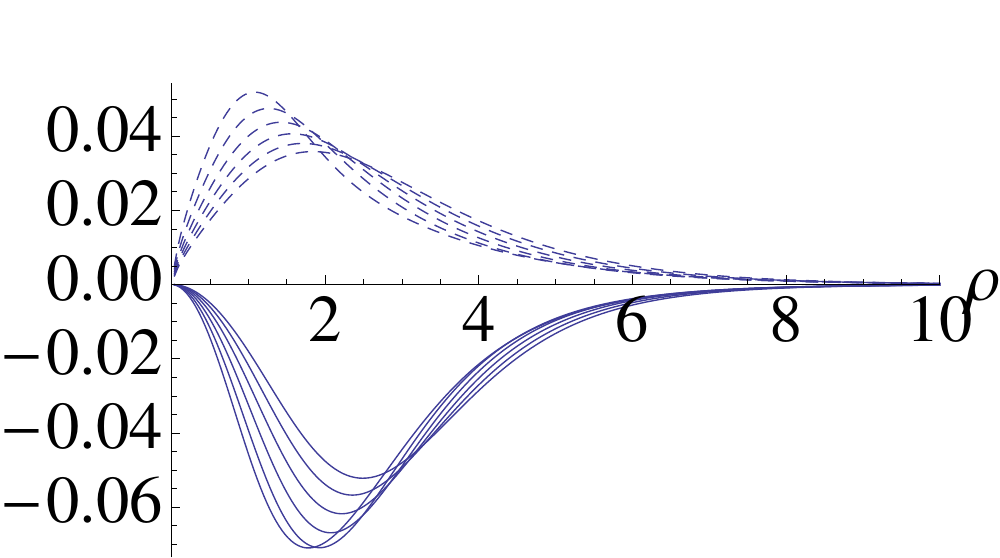}
\end{subfigure}
\begin{subfigure}{.5\textwidth}
\includegraphics[width=0.9\linewidth]{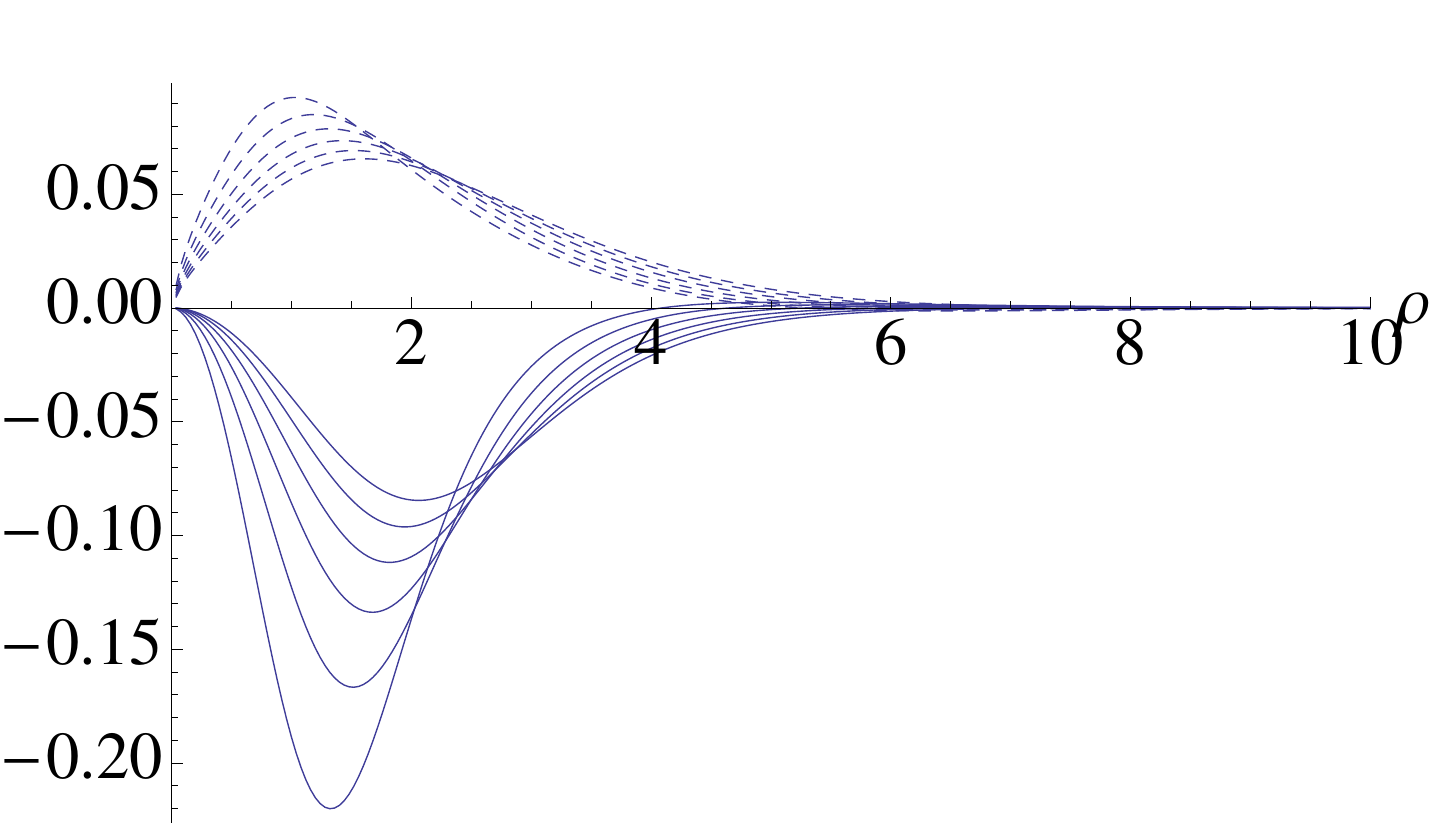}
\end{subfigure}
\caption{Field profiles at $\kappa = 2$, $\kappa_h = 3$, $e=2$, $g = 2$ and $\eta_0/\phi_0=1$ changing $\xi$ between $0.6$ and $1.6$ in steps of $0.2$, with highest $\xi$ corresponding to larger magnitude of magnetic fields. Plots on the second line include electric (dashed line) and magnetic fields (solid line) for both the $A$ (left column) and $G$ (right column) fields.}%
\label{fig4}%
\end{figure}
\begin{figure}[h]
\begin{subfigure}{.5\textwidth}
\centering
\includegraphics[width=0.7\linewidth]{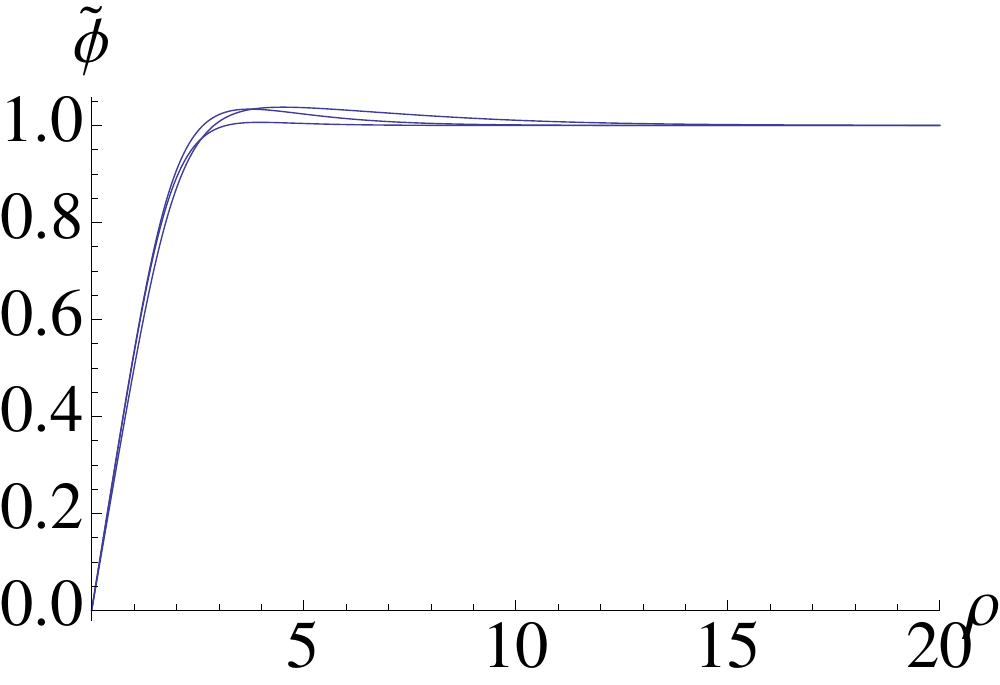}
\end{subfigure}
\begin{subfigure}{.5\textwidth}
\includegraphics[width=0.8\linewidth]{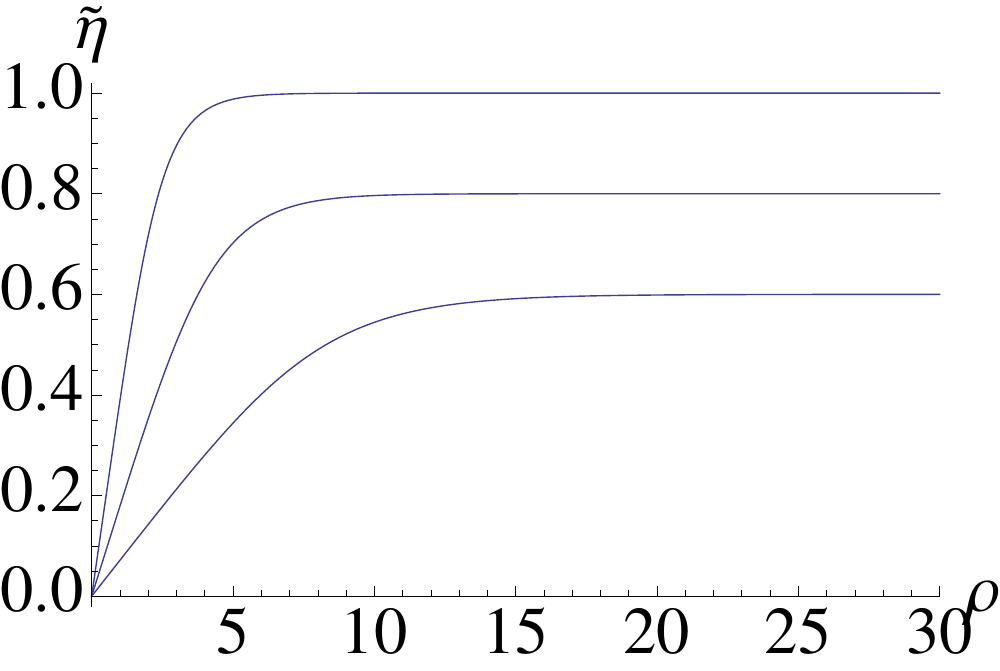}
\end{subfigure}\\
\begin{subfigure}{.5\textwidth}
\centering
\includegraphics[width=0.8\linewidth]{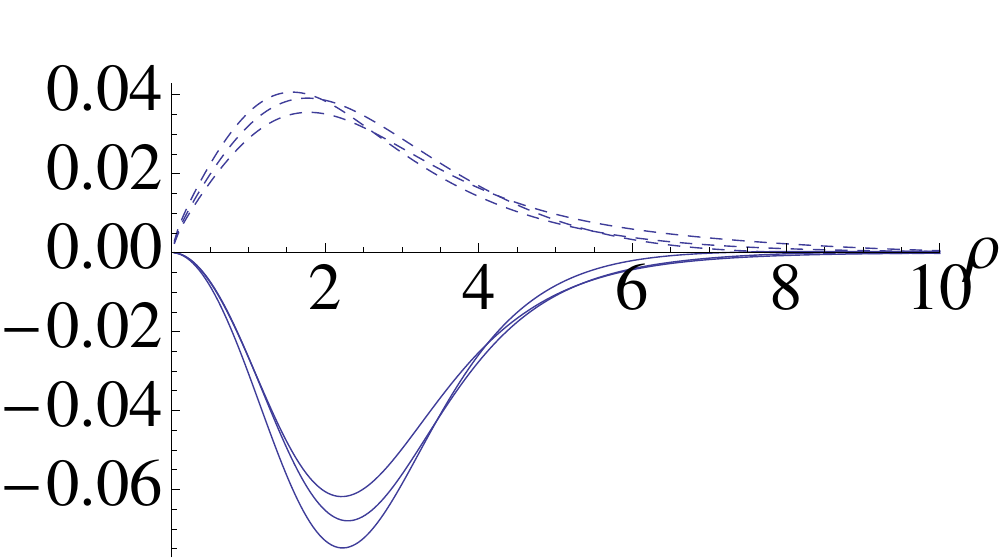}
\end{subfigure}
\begin{subfigure}{.5\textwidth}
\includegraphics[width=0.8\linewidth]{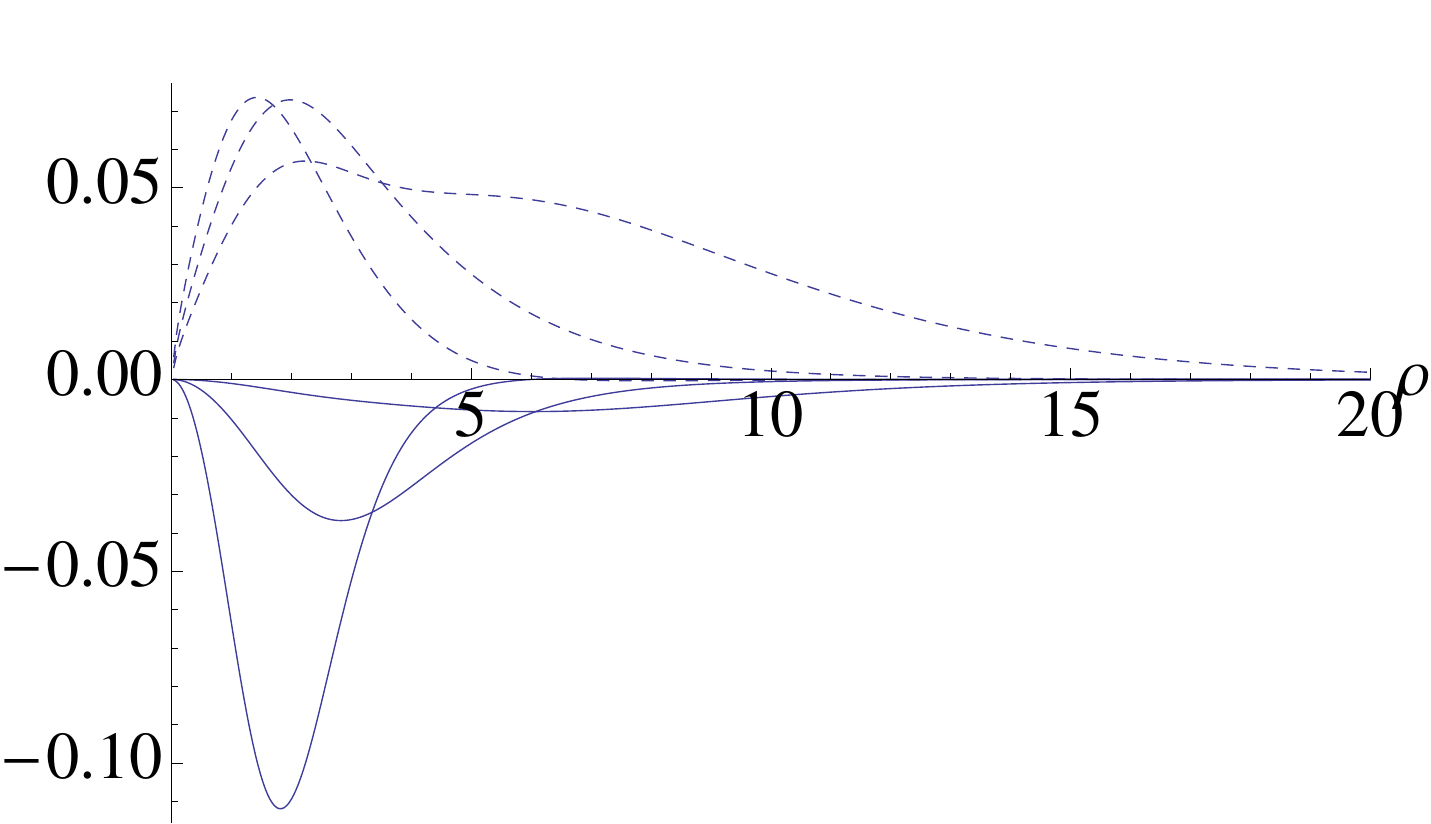}
\end{subfigure}
\caption{Field profiles at $\kappa = 2$, $\kappa_h = 3$, $e=2$, $g = 2$ and $\xi=1$, changing $\eta_0/\phi_0$ between $0.6$ and $1$ in steps of $0.2$, with highest $\eta_0/\phi_0$ corresponding to tighter vortex core. Plots on the second line include electric (dashed line) and magnetic fields (solid line) for both the $A$ (left column) and $G$ (right column) fields.}%
\label{fig4}%
\end{figure}

\end{document}